\documentclass[prb,twocolumn,10pt,aps,showpacs,superscriptaddress,longbibliography,floatfix]{revtex4-1}
\usepackage{graphicx}
\usepackage{amsmath,amssymb,physics}
\usepackage{bm}
\usepackage{dcolumn}
\usepackage{soul} 
\usepackage{psfrag}
\usepackage{float,setspace}
\usepackage{subeqnarray}
	
\usepackage{url}
\usepackage[colorlinks=true,linkcolor=blue,citecolor=blue,urlcolor=blue,breaklinks]{hyperref}

\usepackage{placeins}

\begin{document}

\author{L. Flammia}
\affiliation{Departement Fysica, Universiteit Antwerpen, Groenenborgerlaan 171, B-2020 Antwerpen, Belgium}
\affiliation{School of Science and Technology, Physics Division,
University of Camerino, 62032 Camerino, Italy}
\author{L.-F. Zhang}\email{lingfeng.zhang@uantwerpen.be}
\affiliation{Departement Fysica, Universiteit Antwerpen, Groenenborgerlaan 171, B-2020 Antwerpen, Belgium}
\author{L. Covaci}
\affiliation{Departement Fysica, Universiteit Antwerpen, Groenenborgerlaan 171, B-2020 Antwerpen, Belgium}
\author{A. Perali}
\affiliation{School of Pharmacy, Physics Unit, University of Camerino,
62032 Camerino, Italy}
\author{M. V. Milo\v{s}evi\'{c}}\email{milorad.milosevic@uantwerpen.be}
\affiliation{Departement Fysica, Universiteit Antwerpen, Groenenborgerlaan 171, B-2020 Antwerpen, Belgium}

\title{Superconducting nanoribbon with a constriction: \\a quantum-confined Josephson junction}

\begin{abstract}
Extended defects are known to strongly affect nanoscale superconductors. Here we report the properties of superconducting nanoribbons with a constriction formed between two adjacent step-edges, by solving the Bogoliubov-de Gennes equations self-consistently in the regime where quantum confinement is important. Since the quantum resonances of the superconducting gap in the constricted area are different from the rest of the nanoribbon, such constriction forms a quantum-confined S-S'-S Josephson junction, with a broadly tunable performance depending on the length and width of the constriction with respect to the nanoribbon, and possible gating. These findings provide an intriguing approach to further tailor superconducting quantum devices where Josephson effect is of use.
\end{abstract}


\maketitle

\section{INTRODUCTION}

Nanoscale superconductivity, in which one or more dimensions are smaller than the coherence length,  exhibits a range of interesting phenomena, such as Berezinskii-Kosterlitz-Thouless (BKT) phase transitions \cite{Kos_1973,Kos_1974}, excess conductivity induced by superconducting fluctuations \cite{Asl_1968}, and the superconductor-insulator quantum phase transition at zero temperature \cite{Gol_1998}, to name a few.  In particular, when the size of the superconductor becomes comparable to the electron Fermi wavelength $\lambda_F$, the formation of discretized electronic energy levels results in the oscillations of the density of states at the Fermi level with the size, together with the reconfiguration of the pairing interaction, leading to the oscillatory behavior of superconducting critical temperature $T_c$ and other observables \cite{Bla_1963,Per_1996,Bia_1997,Guo_2004,Oze_2006,Eom_2006,Bru_2009,Qin_2009,Zha_2010,Che_2010,Che_2012,Sha_2015}, i.e. the so-called quantum size effects. 

\begin{figure}[t]
  \centering
  \includegraphics[width=\linewidth]{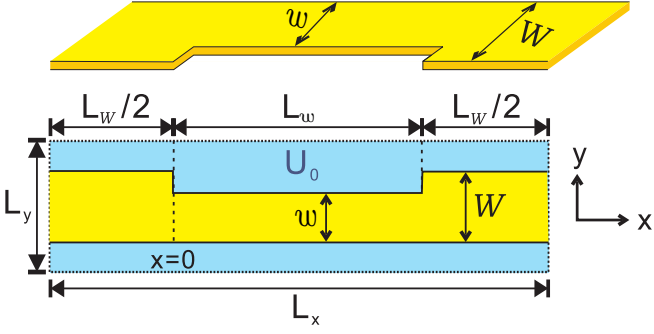}
  \caption{(Color online) Oblique and bird view of a nanoribbon with a constriction formed by step-edges, with indicated geometrical parameters within the computational unit cell.  The width variation of the nanoribbon is realized by imposing a tall potential barrier $U_0$ outside the sample (shaded region). The shown aspect ratio of the simulation region ($L_x:L_y$) greatly under-represents the actual one in the simulations (where $L_x:L_y\approx100:1$).}
  \label{sketch}
\end{figure}

Recent progress in nanotechnology has allowed high-quality superconducting nanostructures to be fabricated with atomic-scale precision \cite{Uch_2017,Pin_2017}.  Superconductivity is realized in atomically thin films even down to a single monolayer.  Quantum size effects have been reported in atomically thin films \cite{Bao_2005}, superconducting nanoparticles \cite{Bos_2010} and islands \cite{Min_2015}, nanowires\cite{Cor_2006} and nanowire arrays \cite{Zha_2016}. However, the low-dimensional superconductivity is strongly influenced by the imperfections such as impurities, disorder and structural defects \cite{Bru_2014}. In nanofilms, increasing disorder results in the phase transition from a superconducting to an insulating state \cite{Gol_1998}. In addition, impurities locally suppress superconductivity, which in superconducting nanowires can promote a phase slip center, giving rise to the broad temperature transition and residual resistance \cite{Lan_1967,McC_1970}. Very recently, a step in atomically thin films was found to have a strong effect on electronic transport \cite{Kim_2016,Zha_2017} and vortex matter \cite{Bru_2014,Yos_2014,Rod_2015}, as a new paradigm in the interplay between the local defects and low-dimensional superconductivity.  As an extended defect, the step does not only scatter electrons (leading to the modification on the overall electronic structure of the sample \cite{Liu_2013}), but also affects the flow of superconducting currents and the proximity-induced superconducting correlations \cite{Kim_2016,Zha_2017}. However, the effect of the lateral step (indentation) in ultrathin yet nanoscale wide superconductors (from here on referred to as nanoribbons) has not been investigated. Such nanoribbons are readily used as building blocks of superconducting quantum devices such as single-photon detectors \cite{Nat_2012}, phase-slip junctions \cite{Moo_2006}, and Josephson junction arrays \cite{Hav_2001}, and can be fabricated even from 2D materials such as graphene, NbSe$_2$, MoS$_2$, WS$_2$, and other \cite{Sen_2017, Ngu_2017, Zhe_2017, Pia_2017}. 
In this paper, we address this issue, with a special attention drawn to superconducting nanoribbons with a constriction formed by two adjacent step-edges (see Fig.~\ref{sketch}). Starting from a long constriction, where two step-edges are far away from each other, we discuss the role of a step-edge and present how it modifies electronic states, the superconducting order parameter and the local density of states (LDOS) in the nanoribbon.  Then, for a short constriction, effectively an extended quantum point contact, we show how the device becomes a quantum-confined Josephson junction, a novel object with quantum-tunable characteristics. Namely, such a device exhibits properties that are governed by quantum size effects, different inside and outside the constriction, hence behaving as a S-S'-S junction with performance broadly dependent on all sizes, temperature, and the Fermi energy (controllable by gating or doping).

The paper is organized as follows. In Sec.~\ref{sec:2}, we present our theoretical approach and details of numerical simulations.  In Sec.~\ref{norm_sec}, we present the normal-state electronic properties for the nanoribbon with a constriction.  Next, we detail the superconducting properties, first for a plain nanoribbon (for necessary background) in Sec.~\ref{sec:4.1}, followed by nanoribbon with a long constriction in Sec.~\ref{sec:4.2}, and finally for the quantum-confined Josephson junction (nanoribbon with a short constriction) in Sec.~\ref{sec:4.3}.  Our findings are summarized in Sec.~\ref{sec:5}.

\section{THEORETICAL MODEL AND NUMERICAL APPROACH}\label{sec:2}

We employ the Bogoliubov-de Gennes (BdG) equations to study the role of step-edges in a superconducting nanoribbon, in which quantum confinement is important. The BdG equations have been successfully used in the past to study the interplay between superconductivity and the quantum confinement, and have revealed many fascinating phenomena - among which the quantum size effect, unconventional vortex states, new Andreev bound states, and quasiparticle interference effect \cite{Sha_2006,Sha_2007,Sha2_2007, Cro_2007}.

The BdG equations are written as:
\begin{equation}
\begin{matrix}
\label{BdG_eq}
\mqty(\hat{K}_0-E_F &\Delta(\vec{r}) \\ \Delta^*(\vec{r})&-\hat{K}_0^*+E_F)\mqty(u_n(\vec{r})\\v_n(\vec{r}))=E_n\mqty(u_n(\vec{r})\\v_n(\vec{r})),
\end{matrix}
\end{equation}
where $u_n(\vec{r})$($v_n(\vec{r})$) are electron(hole)-like wave functions corresponding to the quasiparticle energy $E_n$, $E_F$ is the Fermi energy, and the single-particle Hamiltonian $\hat{K}_0$ reads
\begin{equation}
\hat{K}_0=-\frac{\hbar^2}{2m}\nabla^2+U(\vec{r}),
\end{equation}
with $U(\vec{r})$ the confining potential.
In order to find the quasiparticle energy spectrum $E_n$ and the corresponding wave functions $u_n(\vec{r})$ and $v_n(\vec{r})$, we need to solve the BdG equations self-consistently together with the relation for the pair potential $\Delta(\vec{r})$
\begin{equation}
\label{SC_gap}
\Delta(\vec{r}) = g\sum_{E_n<E_c}u_n(\vec{r})v^*_n(\vec{r})[1-2f(E_n)],
\end{equation}
where $g$ is the coupling constant, $E_c$ is the Debye cutoff energy, and $f(E_n) = [1+\exp(E_n/k_BT)]^{-1}$ is the Fermi distribution function at temperature $T$.  The local density of states is
\begin{equation}
\label{SC_gap1}
N(\vec{r},E) = \sum_n [\delta(E_n-E)\abs{u_n(\vec{r})}^2+\delta(E_n+E)\abs{v_n(\vec{r})}^2].
\end{equation}

We consider a nanoribbon with a dent at the center, as shown in Fig.~\ref{sketch}. The length of the ribbon $L_x$ is much longer than the superconducting coherence length $\xi$, i.e. $L_x  \gg \xi$, with a periodic boundary condition along the ribbon. The dent separates the ribbon into two parts, characterized by different width - the part of length $L_W$ and the width $W$, and the part of length $L_w$ and the width $w$ (being the constriction, i.e. $w\leq W$). The corresponding areas are $S_W=L_W \times W$ and $S_w=L_w \times w$, respectively.  Note that $L_x = L_W+L_w$ and the step-edges are located where the width of the nanoribbon changes. The widths $W$ and $w$ are of the order of the Fermi wavelength $\lambda_F$. Since $\lambda_F<\xi$, $L_x \gg W\,,\,w$.  

Two extreme cases will be taken into consideration.  First we study the role of a single step-edge.  For this purpose, we set $L_W= L_w= L_x/2$ so that the distance between two adjacent step-edges is the farthest. Since $L_x \gg \xi$ we have $L_W\,,\,L_w \gg \xi$, i.e. the interaction between the neighboring step-edges can be neglected.  The other case is the role of a dent where $L_w \approx \lambda_F$. Due to the discrete energy levels inside the dent and the momentum mismatch at the step-edges, the transport properties through the dent will be strongly affected.  In this paper, we do not present the results for the intermediate cases since they can be understood as superposition of two discussed extreme cases.

In order to perform numerical calculations, we embed the nanoribbon in a computational unit cell with area $S = L_x \times L_y$, as shown in Fig.~\ref{sketch}.  The length of the unit cell is the same as that of the ribbon but its width is determined by the condition $L_y > \text{max}\{W,w\}$, so that the single-particle potential barrier
\begin{equation}
U(\vec{r})= \begin{cases}
0 & \text{ outside the ribbon;}  \\ 
U_0 & \text{ inside the ribbon;} 
\end{cases}
\end{equation}
can be applied outside the ribbon to confine the electrons. Since a large magnitude $U_0=20E_F$ is used for the potential barrier, the quasiparticle wave functions $u_n(\vec{r})$ and $v_n(\vec{r})$ decay exponentially at the edges of the ribbon.  

To solve more efficiently the self-consistent BdG equations \eqref{BdG_eq}-\eqref{SC_gap}, we expand $u_n$($v_n$) in terms of the eigenstates $\Psi_l(\vec{r})$ of the single-electron Schr\"odinger equation for the normal state
\begin{equation}\label{Schr}
\hat{K_0}(\vec{r})\Psi_l(\vec{r})=E_l\Psi_l(\vec{r}).
\end{equation}
We first solve Eq.~\eqref{Schr} by expanding $\Psi_l(x,y)$ in terms of plane waves $\phi_{j_x,j_y}(x,y)$, i.e.
\begin{equation}
\label{Four_exp_Psi}
\Psi_l(x,y) = \sum_{j_x,j_y} c^{\,l}_{j_x,j_y} \phi_{j_x,j_y}(x,y),
\end{equation}
where $c^{\,l}_{j_x,j_y}$ are the coefficients for the $l$-th eigenstates and
\begin{equation}
\phi_{j_x,j_y}(x,y) = (L_xL_y)^{-1/2} \text{exp} \Big( i\frac{2\pi j_x}{L_x}x +i\frac{2\pi j_y}{L_y}y \Big),
\end{equation}
with $j_x,j_y \in \mathcal{Z}$.  We define $j=j(j_x,j_y)$.  Then, Eq.~\eqref{Schr} becomes
\begin{equation}
\label{Schr_mat}
T_j c^{\,l}_j + \sum_{j'} U_{jj'} c^{\,l}_{j'} = \zeta_l c^{\,l}_j,
\end{equation}
where
\begin{equation}
T_j = \frac{\hbar^2}{2m} \left[ \left(\frac{2\pi j_x}{L_x}\right)^2 +\left(\frac{2\pi j_y}{L_y}\right)^2\right],
\end{equation}
and
\begin{equation}
U_{jj'}=\int\dd{x}\dd{y}\phi_{j_xj_y}^*(x,y)\,U(x,y)\,\phi_{j'_xj'_y}(x,y).
\end{equation}
Eq.~\eqref{Schr_mat} has a matrix form.  By diagonalizing the relevant matrix, the eigenvalues $\zeta_l$ and eigenfunctions $\Psi_l(\vec{r})$ can be obtained. We remark that $j_x$($j_y$) must remain finite in the numerical calculations, i.e. $j_x = 0,\pm 1,\ldots,\pm j_x^{\text{max}}$ and $j_y = 0,\pm1,\ldots,\pm j_y^{\text{max}}$.  The choice of $j_x^{\text{max}}$ and $j_y^{\text{max}}$ depends on different parameters, including $E_F$, $E_c$, the size of the nanoribbon and the unit cell.  However, when the wave functions are confined in a smaller area, a larger cut-off is needed in order to preserve the same accuracy.  For example, if $w/L_y$ is taken smaller, the number of basis functions associated with the $y$ direction, $j_y^{\text{max}}$, has to be larger. 

Next, we expand $u_n$($v_n$) in terms of $\Psi_l(\vec{r})$ as
\begin{equation}
\binom{u_n(\vec{r})}{v_n(\vec{r})}=\sum_{l}\binom{u^n_l}{v^n_l}\Psi_l(\vec{r}).
\label{Four_exp_uv}
\end{equation}
We use a parameter $\varepsilon$ to control the number of $\Psi_{l}$ in the expansion, such that only those $\Psi_{l}$ with energies $\zeta_l < E_F+\varepsilon E_c$ are included. After inserting Eq.~\eqref{Four_exp_uv} into the BdG Eqs.~\eqref{BdG_eq}, we obtain 
\begin{equation}
    \begin{aligned}
        (\zeta_l-E_F) u^n_l + \sum_{l'} \Delta_{ll'} v^n_{l'} &= E_n u^n_l, \\
        \sum_{l'} (\Delta_{l'l})^* u^n_{l'} + (E_F-\zeta_l) v^n_l &= E_n v^n_l,
    \end{aligned}
\label{BdG_mat}
\end{equation}
where 
\begin{equation}
    \Delta_{ll'} =\int\dd{x}\dd{y}\Psi_{l}^*(x,y)\,\Delta(x,y)\,\Psi_{l'}(x,y),
\end{equation}
and $(\Delta_{l'l})^*$ is the conjugate transpose of $\Delta_{ll'}$.  Similarly to Eq.~\eqref{Schr_mat}, Eq.~\eqref{BdG_mat} has a matrix form as well. The corresponding eigenvalues and eigenstates can be obtained after its diagonalization.

In this paper, we present the results for the following parameters: effective mass $m=2m_e$ ($m_e$ being the electron mass), $E_F = 40~\mathrm{meV}$, $E_c = 24~\mathrm{meV}$, and coupling constant $g$ is set such that the bulk gap at zero temperature is $\Delta_0=1.2~\mathrm{meV}$ and $T_c\approx 8.2~\mathrm{K}$, the coherence length at zero temperature $\xi_0={\hbar}v_F/\left(\pi \Delta_0\right)= 14.7~\mathrm{nm}$ and $k_F\xi_0 = 21$, where $v_F$ is the Fermi velocity and $k_F$ the Fermi wave-vector.  The prototype material can be, e.g., NbSe$_2$ \cite{Gyg_1991, Vir_1999}.  For this set of parameters, we take $L_x=1~\mathrm{\mu m}$, $L_y=12~\mathrm{nm}$.  Then, we find that $j^{\text{max}}_x=500$, $j^{\text{max}}_y=16$ and $\varepsilon=3.5$ yield satisfactory results so that larger cut-off is not necessary.  We also confirm that the features of our results are robust for other $k_F\xi_0$ values so these generic features can be applied to other superconducting materials (e.g., Pb, In, Ga, NbSe$_2$). All the results are calculated at zero temperature, unless specified otherwise.

\section{Normal-state electronic properties}
\label{norm_sec}

\begin{figure}
  \centering
  \includegraphics[width=\linewidth]{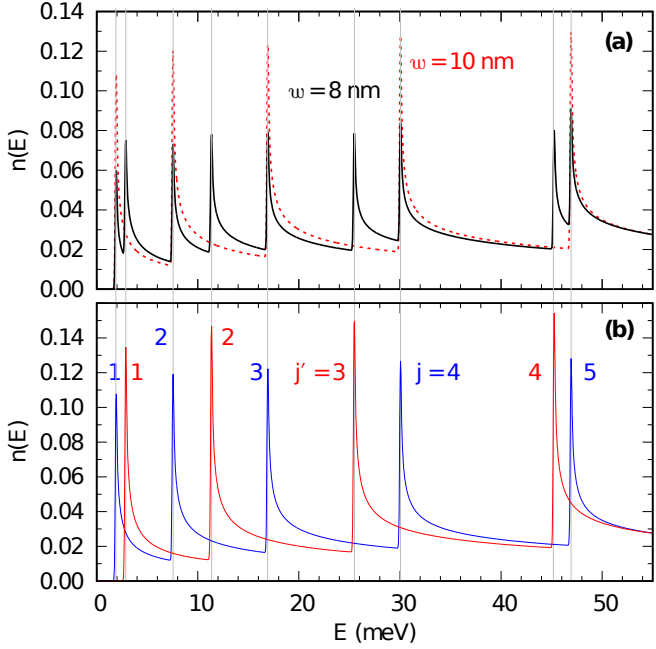}
  \caption{(Color online) The DOS of a nanoribbon with a long constriction, for $W=10~\mathrm{nm}$ and $w=8~\mathrm{nm}$. Panel (a) shows $n(E)$ (solid line), compared to the case with no constriction (dashed line). Panel (b) shows the spatially averaged LDOS over the wider and the constricted part of the nanoribbon,  $n_W(E)$ and $n_w(E)$ respectively, with peaks labelled according to the respective sequence of subbands in two parts of the nanoribbon.}
  \label{FigN1}
\end{figure}

In this section, we examine the normal-state electronic properties of the nanoribbon, since any effect of the constriction on those may further manifest in the superconducting properties.  The normal-state electronic properties can be completely obtained by solving the Schr\"odinger equation \eqref{Schr}.  When there is no constriction ($w=W$), the normal-state electronic structures are well characterized by a series of one-dimensional (1D) subbands, in which the energy dependence of density of states (DOS) of each subband is proportional to $(E-E_j)^{-1/2}$, where $E_j=\frac{1}{m}(\frac{\pi j}{W})^2$ is the threshold energy at the $j$th subband. As a result, the DOS exhibits a peak each time $E_j$ is approached, which corresponds to the van Hove singularity of the standard 1D DOS. An example of this type is shown in Fig.~\ref{FigN1}(a) for the nanoribbon with $W = w= 10~\mathrm{nm}$ (dashed line). The DOS is defined as 
\begin{equation}\label{DOS_N}
n(E) = \sum_l\delta(E-E_l)/(S_W+S_w),
\end{equation}
where $S_W+S_w$ is the area of the ribbon. We use this area because $|\Psi_l(\mathbf{r})|^2$, i.e. the probability density of the wavefunction, is negligible outside of the ribbon due to the very large potential barrier $U_0$.  

A constriction is introduced in the nanoribbon when $W>w$. We initially consider a long constriction where $L_W=L_w=L_x/2 \gg W,w$. In this case we can represent the system as two adjoined nanoribbons of different width. Fig.~\ref{FigN1}(a) shows the corresponding DOS, $n(E)$, of a nanoribbon with $W= 10~\mathrm{nm}$ and $w= 8~\mathrm{nm}$.  It is characterized by the standard 1D DOS, with doubled peaks compared to the plain nanoribbon. The additional peaks in the $n(E)$ can be understood by considering $n_W(E)$ and $n_w(E)$, namely, the spatially averaged local density of states (LDOS) over the $W$-part and the $w$-part of the nanoribbon, respectively. These quantities are calculated as
\begin{equation}
\begin{aligned}
n_W(E) &= \int_W n_r(\mathbf{r},E) d\mathbf{r}/S_W, \\
n_w(E) &= \int_w n_r(\mathbf{r},E) d\mathbf{r}/S_w,
\end{aligned}
\end{equation}
where 
\begin{equation}
    n_r(\mathbf{r},E) = \sum_l|\Psi_l(\mathbf{r})|^2\delta(E-E_l), 
\end{equation}
is the LDOS and $S_W$ ($S_w$) is the area of the $W$-part ($w$-part) of the nanoribbon.  Note that
\begin{equation}\label{n3}
n(E)=[n_W(E)S_W+ n_w(E)S_w]/(S_W+S_w),
\end{equation}
with reference to the definition of $n(E)$ in Eq.~\eqref{DOS_N}. Accordingly, Fig.~\ref{FigN1}(b) shows the individual contribution of $n_W(E)$ and $n_w(E)$ when they are extracted from $n(E)$ [the solid line in Fig.~\ref{FigN1}(a)]. Both $n_W(E)$ and $n_w(E)$ exhibit the standard 1D DOS, as a consequence of the fact both $L_W$ and $L_w$ are sufficiently long. The peaks in $n_W(E)$ are at energies $E_j=\frac{1}{m}(\frac{\pi j}{W})^2$, and those in $n_w(E)$ are at energies $E_{j'}=\frac{1}{m}(\frac{\pi j'}{w})^2$ \footnote{More precisely, the characteristic energies $E_j(W) \rightarrow \frac{1}{m}(\frac{\pi j}{W})^2$ and $E_{j'}(w) \rightarrow \frac{1}{m}(\frac{\pi j'}{w})^2$ when the potential barrier $U_0 \rightarrow \infty$.}.  In short, the DOS of the nanoribbon with a long constriction, which is given by $n(E)$, is featured by the standard 1D DOS with two sets of characteristic energies $E_j(W)$ and $E_{j'}(w)$.   

\begin{figure}
  \centering
  \includegraphics[width=\linewidth]{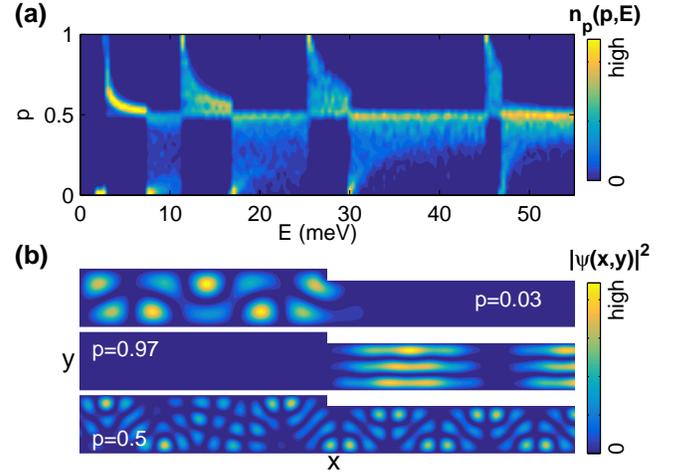}
  \caption{(Color online)  (a) The $p$-resolved DOS, $n_p(p,E)$ (as defined in the text), for the nanoribbon with a long constriction, where $W=10~\mathrm{nm}$ and $w=8~\mathrm{nm}$.  (b) Examples of electronic probability density $|\psi(x,y)|^2$ near $x=0$ for $p=0.03$, $0.97$, and $0.5$.}
  \label{FigN2}
\end{figure}

To understand the obtained behavior of $n_W(E)$ and $n_w(E)$, we study the electronic wavefunctions of the normal state.  For this purpose, we calculate the probability of a wavefunction $\psi(\mathbf{r})$ lying in the constriction, i.e. $p=\int_w|\psi(\mathbf{r})|^2d\mathbf{r}$, and construct the $p$-resolved DOS, $n_p(p,E)$ [see Fig.~\ref{FigN2}(a)], written as
\begin{equation}
n_p(p,E)= \left[\sum_l \delta(E-E_l)\delta(p-p_l)\right]/(S_W+S_w)
\end{equation}
Note that $0 \leqslant p \leqslant 1$ due to the normalization of the wavefunction.  The integral of $n_p(p,E)$ over $p$ is the density of states $n(E)$.  From Fig.~\ref{FigN2}(a), one sees that states mostly lie around $p=0.5$, i.e. when $|\psi(\mathbf{r})|^2$ is spread over the $W$- and $w$-part.  The example of a probability density $|\psi(\mathbf{r})|^2$ for $p=0.5$ is shown in Fig.~\ref{FigN2}(b), and is indeed spread over the entire nanoribbon.

However, we find that there is a large number of states near $p\approx 0$ at $E_j$ and near $p\approx 1$ at $E_{j'}$ [see $n_p(p,E)$ in Fig.~\ref{FigN2}(a)]. States with $p\approx 0$ ($p\approx 1$) are localized in the $W$-part ($w$-part) and decay exponentially in the other part. Examples of these two types of $|\psi(\mathbf{r})|^2$ are also presented in Fig.~\ref{FigN2}(b), with $p=0.03$ and $0.97$, respectively. These two cases of $|\psi(\mathbf{r})|^2$ are in analogy with quantum-well states and are responsible for the standard 1D DOS appearing in $n_W(E)$ and in $n_w(E)$. We also note that there is an exclusion rule between the states with $p>0.5$ and those with $p<0.5$. That is, for the given energy $E$, the states with $p>0.5$ cannot coexist with the state with $p<0.5$, as shown in Fig.~\ref{FigN2}(a). Therefore, the electronic properties in the $W$-part can be very different from those in the $w$-part, especially at the characteristic energies $E_j$ and $E_{j'}$. This property will play a decisive role in the change of superconducting properties at the step-edge(s), where the width of the nanoribbon changes.

\begin{figure}
  \centering
  \includegraphics[width=\linewidth]{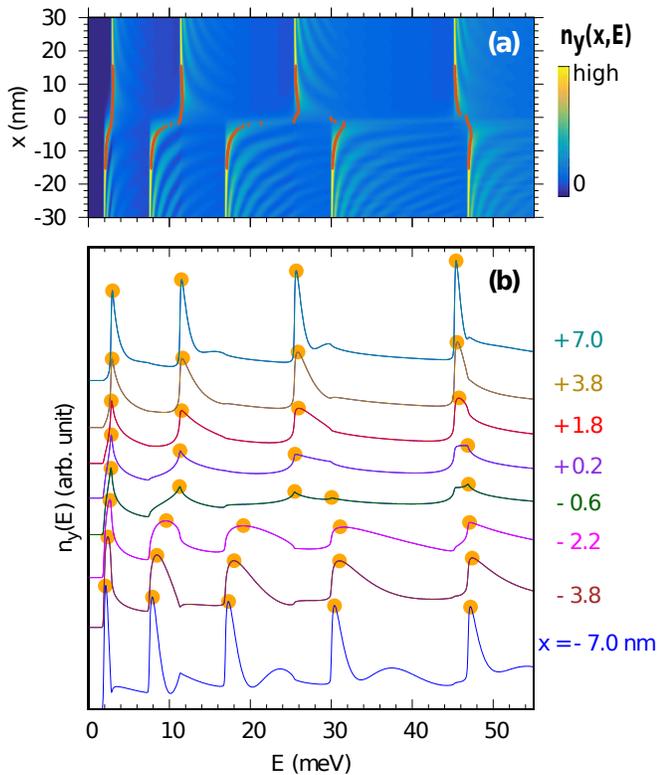}
  \caption{(Color online) (a) The spatial integral of LDOS over $y$, $n_y(x,E)$, near $x=0$ for the nanoribbon with a long constriction, for $W=10~\mathrm{nm}$ and $w=8~\mathrm{nm}$. The evolution of the peaks near $x=0$ is marked by dots. (b) $n_y(E)$ for some selected $x$, vertically displaced for clarity.}
  \label{FigN3}
\end{figure}

How the electronic structure changes near the step-edge (e.g. at $x=0$ in Fig. \ref{sketch}) is an interesting aspect to study. Ref.~\onlinecite{Che_2000} reports that the transitions of the electronic states around the step-edge should be sharp and abrupt. In contrast, Ref.~\onlinecite{Sah_2015} shows that the transition is smooth within a certain lateral extension.  Here we found the abrupt transitions of the electronic states around the step-edge are accompanied by a somewhat smooth transition of the energy of the peaks in LDOS due to the localized states.
To this end, the spatial integral of LDOS over $y$, $n_y(x,E)$, is evaluated near the step-edge, as shown in Fig.~\ref{FigN3}(a). Two opposite behaviors are displayed in the vicinity and far from the step. In the former case, peaks are gradually shifted in energy, while they are significant and $x$-independent in the latter case, appearing at typical energies $E_j$ ($E_{j^\prime}$) in the $W$-part ($w$-part).

Near the step, broad and low peaks evolve by changing their position in energy. Bearing in mind that electronic states are energetically well defined, modifications of the position and the shape of the peaks can be inferred by noting that wavefunctions exponentially decay when passing through the step. The occurrence and the shape of a peak at energies $E_j$ ($E_{j^\prime}$) depends on how much the wavefunction is spread over the $W$-part ($w$-part) around the step. The larger the distance is from the step, the sharper the peak is because the wavefunction is well localized on that part. When the distance from the step is progressively reduced, the resulting position of a peak can be shifted in energy since a superposition effect may occur.

\begin{figure}
  \centering
  \includegraphics[width=\linewidth]{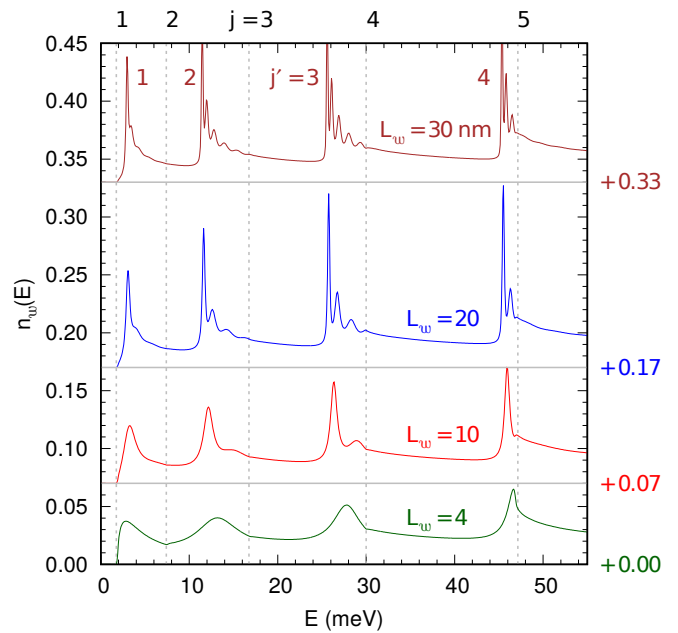}
  \caption{(Color online) $n_w(E)$ of nanoribbons with $W=10~\mathrm{nm}$ and $w=8~\mathrm{nm}$, for different $L_w$. The lineplots of $n_w(E)$ are vertically displaced for clarity. Dashed vertical lines indicate $E_j$, the positions of peaks in $n_W(E)$.}
  \label{FigN4}
\end{figure}

After understanding the properties of the nanoribbon with a long constriction, we turn focus to the case of a short constriction ($L_w\sim w$).  In this case, the length of the $W$-part, $L_W$, is always kept sufficiently long so that the electronic properties in the $W$-part are independent of both $L_W$ and $L_w$, i.e. $n_W$ always exhibits the standard 1D DOS as shown in Fig.~\ref{FigN1}(b).
On the other hand, different electronic properties may occur when $L_w\approx w$ because, in contrast to the long constriction, here the short constriction can be viewed as an extended quantum point contact.

To demonstrate how the electronic properties change with decreasing $L_w$, we show in Fig.~\ref{FigN4} the $n_w(E)$ dependence for the nanoribbon with $W=10~\mathrm{nm}$ and $w=8$ nm, for different lengths of the constriction ($L_w$). For $L_w=30$ nm, $n_w(E)$ still exhibits characteristics of standard 1D DOS, but the appearance of the main peak at $E_{j^\prime}$ is accompanied by several secondary peaks, due to the discrete energy levels induced in the constriction by the quantum confinement in the $x$ direction. As $L_w$ is decreased, the number of secondary peaks decreases and the energy spacing between those peaks becomes larger.  For example, two secondary peaks appear after the $j^\prime=3$rd peak for $L_w=20$ nm, while only one remains for $L_w=10$ nm. Meanwhile, the main peaks are displaced in energy from $E_{j^\prime}$, towards the closest peak of $n_W(E)$ at $E_j$.  These shifts can be larger for smaller values of $L_w$ [see in particular the case of $L_w=4$ nm in Fig.~\ref{FigN4}].  In addition, peaks of $n_w(E)$ become more pronounced as approaching $E_j$ [cf. for example the $j^\prime=4$th peak to the preceding peaks for $L_w=4$ nm]. 

It is worth mentioning that for short $L_w$, $n(E) \rightarrow n_W(E)$, following from Eq.~\eqref{n3} due to $S_W \gg S_w$.  Moreover, we find that all the electronic states are mixed and spread across the entire nanoribbon. Therefore, there are no localized states in the short constriction, in contrast to the long one. 

\begin{figure}
  \centering
  \includegraphics[width=\linewidth]{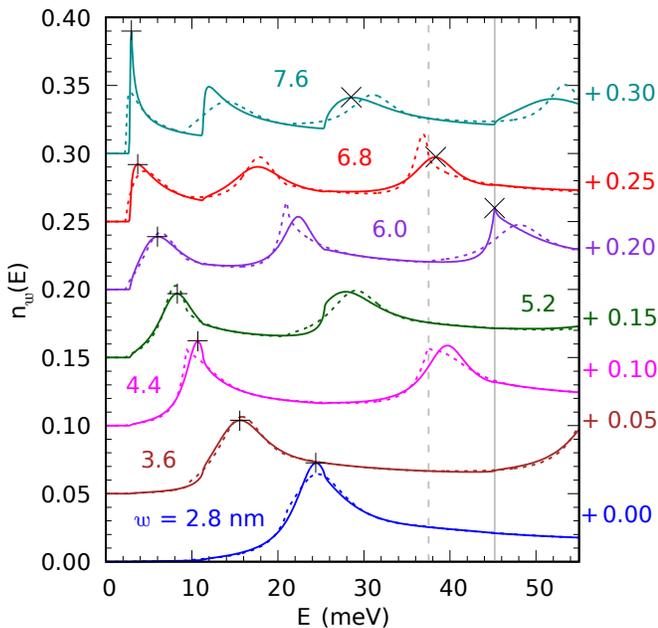}
  \caption{(Color online) The $n_w(E)$ characteristics in the case of short constriction ($L_w=4$ nm), for different width of the constriction ($w$, indicated in the figure) and fixed width of the nanoribbon [either $W=8$ nm (solid lines) or $W=8.8$ nm (dashed lines)]. The series of $n_w(E)$ are vertically displaced for clarity. $+$ and $\times$ indicate the first and the third peak, respectively. Vertical lines indicate the nearest peak in the corresponding $n_W(E)$ [for $W=8~\mathrm{nm}$ (solid) and $W=8.8~\mathrm{nm}$ (dashed)].}
  \label{FigN5}
\end{figure}

Finally, Fig.~\ref{FigN5} shows a series of $n_w(E)$ for the shortest considered constriction $L_w=4~\mathrm{nm}$, now for different widths of the constriction ($w$). The solid lines and the dashed lines represent the case of nanoribbons with $W=8$ nm and $W=8.8$ nm, respectively. In all cases the $n_w(E)$ characteristic exhibits a series of broad and smooth peaks, whose shape depends on whether peaks of $n_W$ at $E_j$ are close to them.  These peaks are nearly independent of the width of the nanoribbon $W$, and shift to higher energy with decreasing the width of the constriction $w$. Therefore, the case of a short constriction case can also be viewed as a nanoribbon coupled with a spectrally broadened quantum dot.

\section{Superconducting properties}\label{sec:4}

\subsection{Background: superconducting nanoribbons without a constriction}\label{sec:4.1}

\begin{figure}
  \centering
  \includegraphics[width=\linewidth]{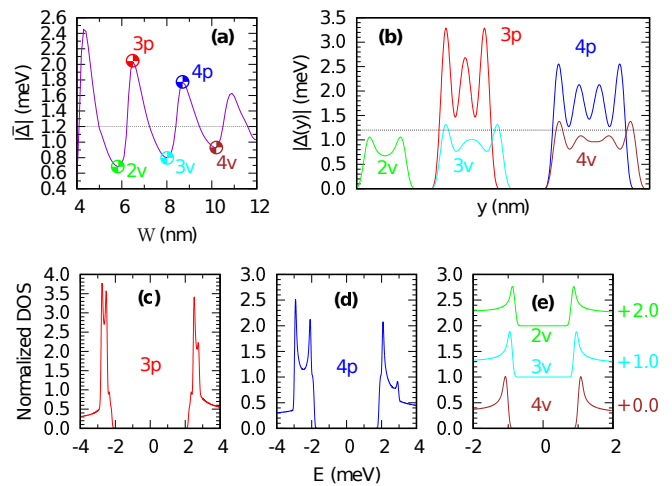}
  \caption{(Color online) Superconducting properties of a homogeneous nanoribbon ($w=W$). Panel (a) shows the spatially averaged order parameter, $|\bar{\Delta}|$, as a function of width $W$. The characteristic behavior of $|\bar{\Delta}|$ is identified at widths indicated by dots, labeled according to the order of the oscillations along the $y$ direction and with the letter indicating the peak ($p$) or valley ($v$) of $|\bar{\Delta}(W)|$. Dotted line shows $\Delta_0$ in the bulk limit ($W \rightarrow \infty$). Panel (b) shows the spatial profile of the order parameter ($|\Delta(y)|$), for the five cases indicated in (a). Panels (c) - (e) show the corresponding DOS as a function of energy $E$. Note that DOS profiles in panel (e) are shifted vertically for clarity.}
  \label{SC0}
\end{figure}

After comprehending the fundamental normal-state properties of the nanoribbon with a constriction, we move on to the analysis of the superconducting state.
The superconductivity in a nanoribbon with no constriction (i.e. $w=W$) has already been studied in detail elsewhere \cite{Sha_2006,Cro_2006}. Here, we repeat some relevant properties of the superconducting nanoribbon under quantum confinement, which will be used as a reference later on when considering the nanoribbon with a constriction ($W>w$).  

Fig.~\ref{SC0}(a) shows the spatially-averaged superconducting order parameter, $\bar{\Delta}$, as a function of the width $W$ of the nanoribbon. It exhibits quantum size oscillations as a function of the width, due to the fact that the normal-state single-electron band splits into a series of subbands under the quantum confinement effect. These subbands shift in energy with $W$, giving rise to the variations in the DOS at $E_F$, i.e. the number of electrons which can contribute to Cooper-pairing.  As $W$ is varied, when the bottom of a new subband approaches $E_F$, the DOS increases together with a substantial reconfiguration of the pairing interaction, leading to the resonant enhancement of superconductivity.  

The quantum-confinement regime for the transverse direction of the electron motion results in the spatial variations of the order parameter along the $y$ direction, in reference to the sketch of the system in Fig.~\ref{sketch}. $\Delta(y)$ is shown in Fig.~\ref{SC0}(b) for characteristic five cases in Fig.~\ref{SC0}(a), i.e. two for resonance cases $3p$ and $4p$ and three for off-resonance cases $2v$, $3v$, and $4v$ [the number in these labels indicates the order of oscillations along the $y$ direction, and $p$ ($v$) stands for peak (valley) in $\bar{\Delta}(W)$]. $\Delta(y)$ of the resonance cases is stronger in amplitude and more spatially inhomogeneous than in off-resonance cases.  

Due to the pronounced inhomogeneity of the order parameter under the resonance condition, a multi-gap structure can form in the DOS [see Figs.~\ref{SC0}(c) and (d)] \cite{Che_2010}, detectable in experimentally measured tunneling spectrum. In addition, new type of Andreev reflection and Tomasch oscillations are also induced due to strongly inhomogeneous order parameter.  In contrast, when the bottom of any present subband is away from $E_F$, the superconductivity is in the off-resonant condition where the corresponding DOS is characterized by a conventional BCS gap structure [as in Fig.~\ref{SC0}(e)].

\subsection{Superconducting nanoribbons with a long constriction}\label{sec:4.2}

\begin{table}
\setlength{\tabcolsep}{10pt}
\begin{tabular}{ | c | c | c |}
 case & conditions for $W$/$w$ side & ($W$,$w$)[$\mathrm{nm}$] \\ [.5ex]
 I &  RES/RES & $8.8$, $6.5$ \\  [.3ex]
 II &  RES/OFF-RES & $8.8$, $8$ \\  [.3ex]
 III &  OFF-RES/RES & $10.1$, $6.5$ \\ [.3ex]
 IV &  OFF-RES/OFF-RES & $10.1$, $5.8$  
\end{tabular}
\caption{The conditions for the characteristic cases I-IV for a nanoribbon with a long constriction, and the corresponding widths $W$ and $w$.  The resonance (RES) and off-resonance (OFF-RES) conditioning corresponds to peaks and valleys indicated in Fig.~\ref{SC0}(a), respectively.}
\label{table1}
\end{table}

Next we consider the superconducting state for a sample with a constriction, thus for $w < W$.  First, we study the influence of a single step-edge, i.e. $L_W$, $L_w \gg \xi$, where the interaction between the adjacent step-edges is negligible. It is clear that the superconducting properties far away from the step-edges are the same as those of plain nanoribbons with the corresponding width (either $W$ or $w$). However, the superconducting features are essentially different depending on the resonance or off-resonance configuration selected for the pair $(W,w)$. Thus, we present the results for the four possible cases, whose parameters are given in Table~\ref{table1}.

\begin{figure}
  \centering
      \includegraphics[width=1\linewidth]{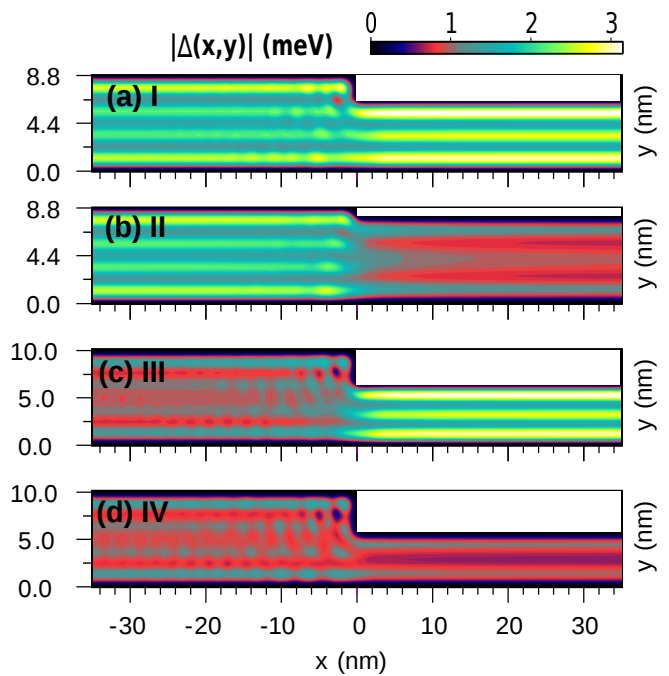}
  \caption{(Color online) The spatial distribution of the order parameter, $\abs{\Delta(x,y)}$, near the step-edge for the cases I-IV of Table \ref{table1}.}
    \label{OPxy}
\end{figure}

\begin{figure}
  \centering
  \includegraphics[width=1\linewidth]{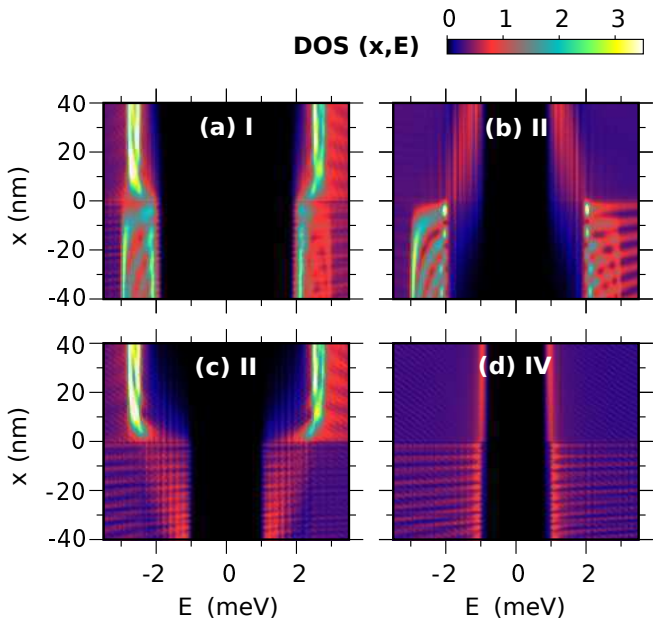}
  \caption{(Color online) LDOS averaged across the width of the sample, DOS($x$,$E$), for the selected cases I-IV of Table~\ref{table1}, plotted in the vicinity of the step-edge.}
  \label{DOSx}
\end{figure}

Fig.~\ref{OPxy} shows the spatial distribution of the order parameter, $\abs{\Delta(x,y)}$, for the characteristic four cases described in Table~\ref{table1}. The corresponding LDOS, averaged over width of the nanoribbon [DOS($x$,$E$)], is shown in Fig.~\ref{DOSx}.  We only present the results in the vicinity of the left step-edge (with situation at the other step-edge being same, i.e. mirror-symmetric). We find that the behaviors of the order parameter and DOS are well described by the normal-state electronic structures in all cases.  In case IV, off-resonant superconductivity is present in both parts of the sample (i.e. with widths $W$ and $w$). The corresponding $E_F$ is away from the bottom of every subband at energies $E_j$ and $E_{j^\prime}$, and the Cooper pairs, formed by the electronic states with $p\approx0.5$, are dominant. Since these electronic states spread over entire nanoribbon, the superconducting properties do not show significant variations when crossing the step-edge.  As shown in Fig.~\ref{OPxy}, the order parameter of case IV does not have a sharp change in the vicinity of the step-edge, differently from cases II and III. In addition, the DOS [Fig.~\ref{DOSx}(d)] exhibits conventional BCS gap in both the constriction and the rest of the nanoribbon.

In case III, the superconducting state is in the resonance configuration in the constricted ($w$-part), and in the off-resonance configuration in the rest of the nanoribbon ($W$-part). Thus, $E_F$ is near $E_{j^\prime}$ but far away from $E_j$ so that the normal-state electronic states with $p\approx1$ are dominant over the $w$-part. 
As a result, the superconducting properties on one side are very different from those on the other side of the step-edge. For example, as shown in Fig.~\ref{OPxy}(c), the order parameter drops dramatically when crossing the step-edge from the narrower $w$-side to the $W$-side. Generally, the spatial variation of the order parameter is defined by its characteristic length, i.e. the coherence length $\xi$, inside the vortex core and at the S-S' interfaces.  However, the enhancement of the order parameter in the constricted $w$-part is here induced by the normal-state electronic states with $p\approx1$. These states decay exponentially when crossing the step-edge, and the characteristic length scale is of the order of $\lambda_F$ ($\xi\approx 10-1000\lambda_F$ in conventional superconductors \cite{Alt_2013}).  Therefore, the order parameter exhibits a fast variation within distance of the order of $\lambda_F$ at the step-edge, in a similar fashion to the occurrence of Friedel-like oscillations near the surface of a superconductor.  Due to this feature, a superconducting nano-structure with a step-edge behaves as a rather sharp, ideally contacted S-S' junction.

The localization of the resonant superconducting properties in the long constriction can also be seen in the DOS($x$,$E$) for case III [Fig.~\ref{DOSx}(c)]. As discussed above, the superconducting gap is larger and the coherence peaks are more pronounced in the $w$-part than in the $W$-part.  However, these features suddenly disappear at the step-edge leading to a dramatic change in the gap structures.  This is also due to the normal-state electronic states with $p\approx1$ which decay at the step-edge.  Note that, due to the large variation of the gap amplitude near the step-edge, the inverse proximity effect is clearly visible.  Its magnitude varies slowly away from the step-edge because the length scale of the variation is related to $\xi$.  However, this effect is much less significant when compared to the effects related to the abrupt change in the normal-state electronic states in the vicinity of the step-edge.

The case II is inversely analogous to the case III, with superconducting state being in resonance in the $W$-part, while off-resonant superconductivity is present in the constriction ($w$-part). Therefore, the same conclusions can be deduced as done in case III, but for opposite sides of the step-edge.  The change in superconductivity is dramatic when the step-edge is crossed, and this variation can be seen both in the profile of the order parameter [Fig.~\ref{OPxy}(b)] and the DOS [Fig.~\ref{DOSx}(b)]. 

Case I is peculiar because resonant superconductivity is attributed to both sides of the step-edge, indicating that $E_F$ approaches both $E_{j^\prime}$ and $E_j$. The variation of the order parameter near the step-edge is not large, being similar to case IV [cf. Figs.~\ref{OPxy}(a) and (d)].  However, the resonance conditions in the $W$-part and $w$-part are induced by the normal-state electronic states with $p\approx0$ and with $p\approx1$, respectively. Both of them are localized and the corresponding probability density decays exponentially at the step-edge so that the superconducting electronic structures abruptly change when the step is crossed. As shown in Fig.~\ref{DOSx}(a), the multi-gap features in the $W$-part are different from those in the $w$-part, but they all coalesce into single gap near the step-edge. In fact, coherence peaks are strongly suppressed at the step-edge, as a sign of the loss of the superconducting coherence. Therefore, the superconducting properties in the resonance configuration are more sensitive to the imperfections such as impurities, disorder, surface roughness and structural defects because of the localization of the electronic states, leading to the suppression of the superconducting coherence at the imperfections. In this case, the critical current is limited by the weakest point of the nanoribbon.  

\subsection{Superconducting nanoribbons with a short constriction}\label{sec:4.3}

\begin{figure}
  \centering
  \includegraphics[width=\linewidth]{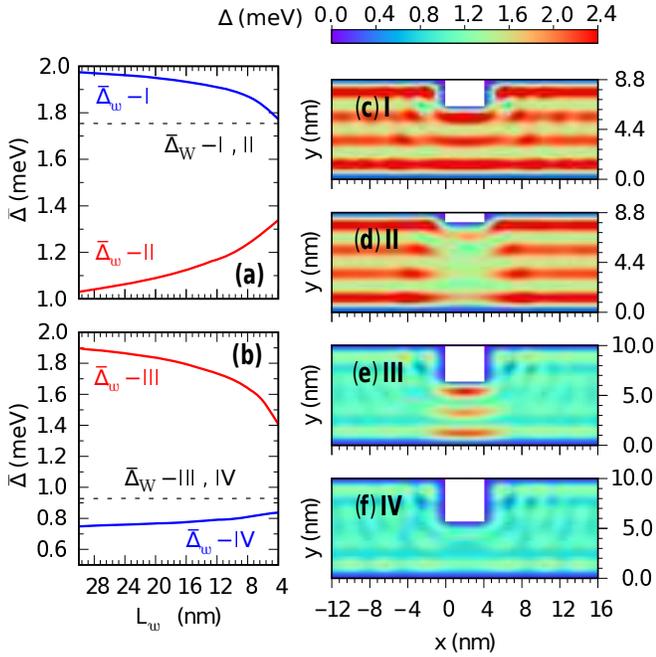}
  \caption{(Color online) The spatially averaged order parameter in and out of the constriction, $\bar{\Delta}_w$ and $\bar{\Delta}_W$ respectively, as a function of $L_w$, for the cases I and II in panel (a) and cases III and IV in panel (b).  Panels (c)-(f) show the corresponding $\Delta(x,y)$ for $L_w=4~\mathrm{nm}$.}
  \label{FigSs1}
\end{figure}

As previously mentioned, a nanoribbon with a long constriction can be viewed as two adjoined nanoribbons, each with different width. Such picture is no longer valid when the constriction is sufficiently short. Namely, when $L_w$ is comparable to the superconducting coherence length $\xi$, the proximity effect plays an important role, reducing the difference in superconducting gap between the constriction and the rest of the nanoribbon. In Fig.~\ref{FigSs1}(a,b), we show spatial averages $\bar{\Delta}_w$ inside and $\bar{\Delta}_W$ outside the constriction as a function of the length of the constriction $L_w$, for the characteristic cases I-IV of Table~\ref{table1}. We note that $\bar{\Delta}_W$ is independent of $L_w$ in all cases, while $\bar{\Delta}_w$ approaches $\bar{\Delta}_W$ with decreasing $L_w$. The corresponding spatial profiles $\Delta(x,y)$ near the constriction for the cases I-IV and $L_w=4$ nm are presented in Fig.~\ref{FigSs1}(c)-(f), respectively. In this limit of short constriction, $L_w$ is comparable to the Fermi wavelength ($L_w\approx\lambda_F$) so that the superconducting gap difference at the step-edge strongly diminishes [see e.g. cases I and IV in Fig.~\ref{FigSs1}(c) and (f), respectively]. 
However, the spatial arrangement of $\Delta(x,y)$ is still consistent with the selected cases I-IV for the nanoribbon with a long constriction [cf. Figs.~\ref{OPxy}(a)-(d)]. 

\begin{figure}
  \centering
  \includegraphics[width=\linewidth]{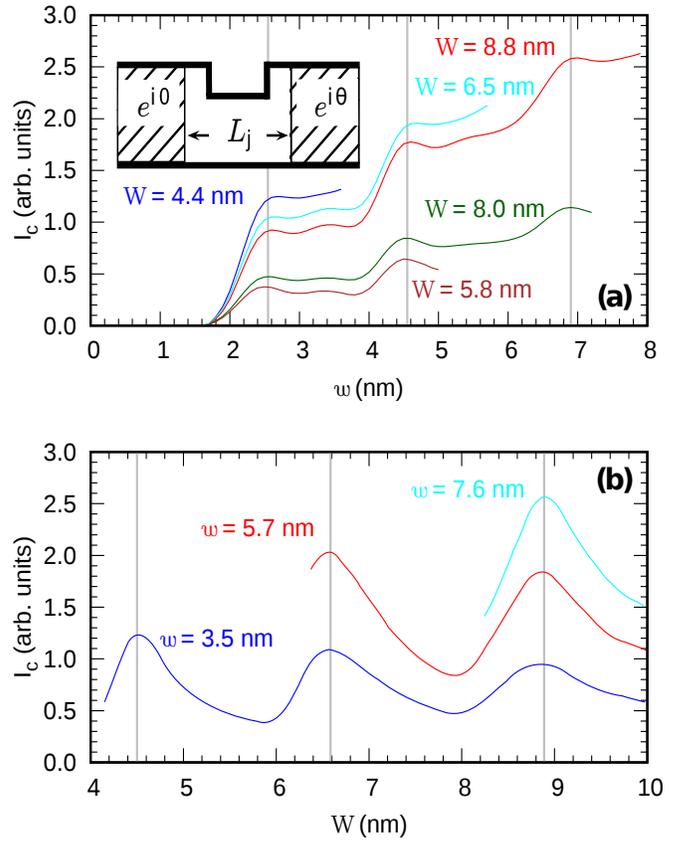}
  \caption{(Color online) (a) The critical current $I_c$ of the quantum-confined Josephson junction as a function of the width of a short constriction $w$, for different width $W$ of the nanoribbon, length of the constriction $L_w=4$ nm, and considered length of the junction $L_j=7$ nm. A sketch of the system is shown in the inset. (b) $I_c(W)$ for different widths of the constriction $w$, and other parameters same as in (a). Peaks are emphasized by vertical lines. Results are limited to range $w<W$, as governed by the geometry of the considered system.}
  \label{FigSs2}
\end{figure}

The short constriction can thus be viewed as a quantum point contact, which results in a point-contact Josephson junction in the nanoribbon. To analyze its transport properties, we calculate the Josephson current passing through the short constriction. For this purpose, we set up a junction link of length $L_j=7$ nm around the short constriction of length $L_w=4~\mathrm{nm}$, as shown in the inset of Fig~\ref{FigSs2}(a). Inside the link, the superconducting gap $\Delta$ is calculated self-consistently. Outside the link, we fix the phase of the order parameter as $\Delta(x<-1.5~\mathrm{nm})=|\Delta|e^{i0}$ and $\Delta(x>5.5~\mathrm{nm})=|\Delta|e^{i\theta}$, such that a phase difference $\theta$ is imposed between the two sides of the link. Then, the supercurrent density is calculated as 
\begin{align*}
\label{densityJ}
\vec{J}(\vec{r}) &= \frac{e\hbar}{2mi} \sum_{E_n<E_c}\left \{ f(E_n)u_n^*(\vec{r})\nabla u_n(\vec{r}) \right.\\
&+ \left. [1-f(E_n)]v_n(\vec{r}) \nabla v_n^*(\vec{r})-\text{h.c.}\right \},
\end{align*}
and satisfies the continuity condition $\nabla \cdot \vec{J} = 0$ inside the link due to the self-consistent $\Delta$ \cite{Spu_2010,Cov_2006}, resulting in the current conservation inside the link [i.e. $I(x) \equiv I = \int\dd{y} J_x(x,y)$].  Outside the link, $\vec{J}$ is discontinued due to the fixed phase of the order parameter, but these regions are simply treated as current sources in the present approximation.

The calculated critical current $I_c$ of the junction exhibits a step-like variation as a function of the width of the constriction $w$ for different values of $W$, as shown in Fig.~\ref{FigSs2}(a). Steps in $I_c$ occur each time the Fermi energy $E_F$ is crossed by a peak of $n_w$ or, in an equivalent formulation, when a new channel of conductance takes part in the current transport. This step-like behavior bears similarities with the quantum conductance in the SNS junction \cite{Fur_1991}. By changing the width of the ribbon $W$, steps of $I_c$ occur for same $w$ because the peaks of $n_w$ are nearly independent of $W$.

Moreover, $I_c$ has large value when the nanoribbon is in the resonance condition, i.e. for $W=8.8$, $6.5$ and $4.4$ nm, in contrast to the low $I_c$ for $W=8.0$ and $5.8$ nm, when nanoribbon is in the off-resonant condition [see Fig.~\ref{FigSs2}(a)]. It is worth noting that $I_c$ for $W=4.4$ nm is even more enhanced than the one for $W=8.8$ nm.  Such behavior is more clearly seen in Fig.~\ref{FigSs2}(b), where the critical current as a function of $W$, $I_c(W)$, is reported.  The $I_c(W)$ characteristic exhibits the quantum-size oscillations with increasing amplitude as $W$ is smaller. This is due to the fact that $\bar{\Delta}(W)$ is more enhanced at resonance in narrower nanoribbons, as shown in Fig.~\ref{SC0}(a). Thus, for fixed $w$, high $I_c$ is obtained for smallest $W$ that corresponds to a resonance condition. Fig.~\ref{FigSs2}(b) also implies that $I_c(W)$ converges for $W \rightarrow \infty$ so that it only depends on $w$ in this limit.

\begin{figure}
  \centering
  \includegraphics[width=\linewidth]{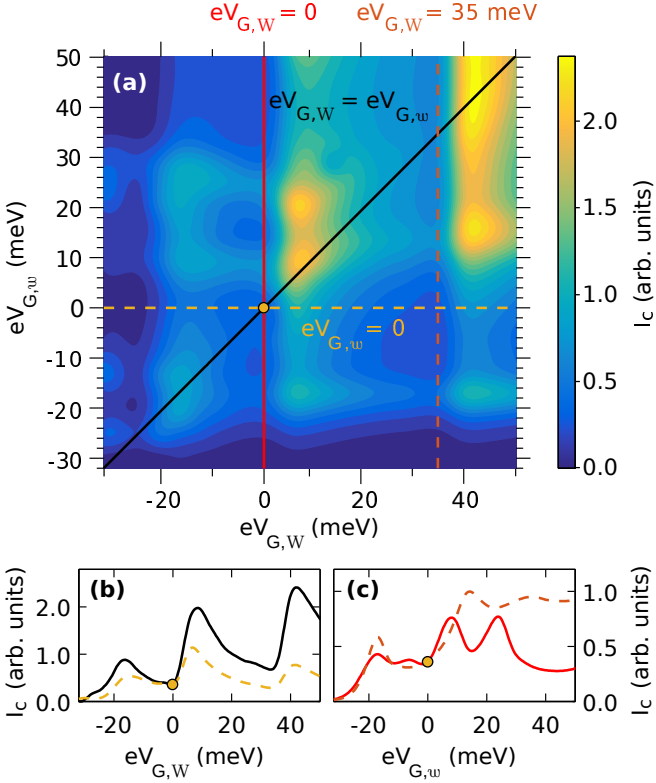}
  \caption{(Color online) (a) Critical current $I_c$ of the quantum-confined Josephson junction made as a constriction of width $w=4$ nm and length $L_w=4$ nm inside a nanoribbon of width $W=6$ nm, as a function of the electronic potential energies $eV_{G,W}$ and $eV_{G,w}$ stemming from gate voltages respectively applied outside the junction ($W$-part) and inside the junction ($w$-part).  Zero gating voltages (marked by open dot) correspond to the reference Fermi energy of $E_F=40$ meV. The profiles of $I_c(eV_{G,W})$ along $eV_{G,W}=eV_{G,w}$ (same gating in entire sample) and $eV_{G,w}=0$ (no gating in the junction) are plotted in panel (b), with line types corresponding to those shown in (a).  The profiles of $I_c(eV_{G,w})$ for $eV_{G,W}=0$ (no gating outside the junction) and $eV_{G,W}=35$ meV are plotted in panel (c), with line types corresponding to those shown in (a).}
  \label{FigSs3}
\end{figure}
\begin{figure}
  \centering
  \includegraphics[width=\linewidth]{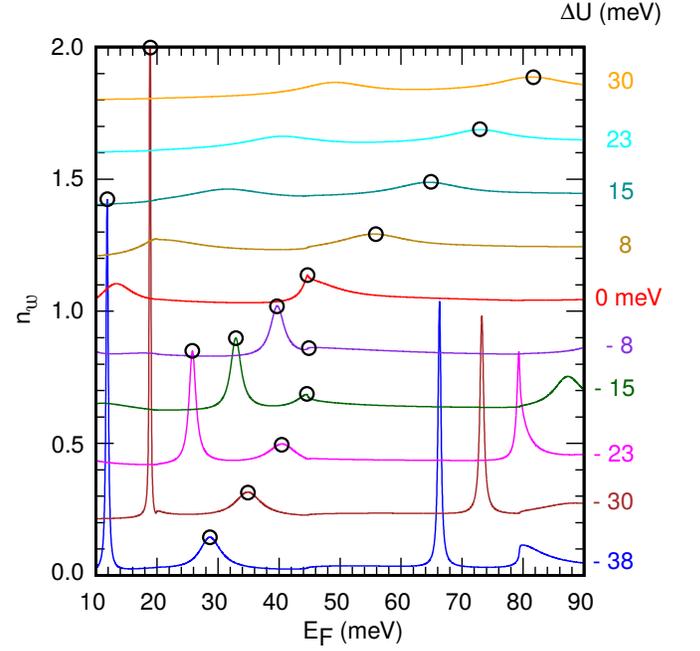}
  \caption{(Color online) The normal state LDOS spatially averaged over the junction, $n_w(E_F)$, for given gating difference $\Delta U=eV_{G,W}-eV_{G,w}$ between the nanoribbon and the junction, for the sample with $W=6$ nm, $w=4$ nm and $L_w=4$ nm.  Open circles highlight the shift of the peaks with $\Delta U$.  Note that additional peaks occur when $\Delta U<0$, i.e. for potential well realized inside the junction.}
  \label{FigSs4}
\end{figure}

Finally, we show that the critical current $I_c$ of a short constriction can also be broadly tuned by electronic gating. Concretely, we apply gate voltages $V_{G,W}$ outside the junction (in the $W$-part) and $V_{G,w}$ inside the junction (in the $w$-part), respectively.  The gate voltage $V_G$ is assumed to induce a chemical potential shift in the gated part of the sample, by electronic potential energy $eV_G$ \cite{Zhe_2017, Pia_2017}, with $e$ being the electron charge.  The evolution of the $I_c$ of the quantum-confined junction as a function of gate voltages $V_{G,W}$ and $V_{G,w}$ is shown in Fig.~\ref{FigSs3}(a), for the short constricted nanoribbon with $W=6$ nm, $w=4$ nm, and $L_w=4$ nm.  To give a better understanding of the features of $I_c$ under applied gate voltages, we also present the profile of $I_c(eV_{G,W})$ for $eV_{G,W}=eV_{G,w}$ (same gating in entire sample) and for $eV_{G,w}=0$ (no gating in the junction) in Fig.~\ref{FigSs3}(b), and the profile of the $I_c(eV_{G,w})$ for $eV_{G,W}=0$ (gating only in the junction) and for $35$ meV (fixed gating outside the junction) in Fig.~\ref{FigSs3}(c).  No applied gating [indicated by open dot in Figs.~\ref{FigSs3}(a)-(c)] corresponds to the reference sample with $E_F=40$ meV.

As a general trend, $I_c$ increases under positive gate voltages due to the introduction of more charge carriers and more channels of conductance taking part in the current transport.  On the other hand, the negative gate voltages reduce the $I_c$ to zero, with the complete depletion of charge carriers reached for $eV_{G,W}$ or $eV_{G,w}$ below $-E_F$.

We note that $I_c$ can be tuned by either voltage $V_{G,W}$ or $V_{G,w}$. In the former case, as seen from Fig.~\ref{FigSs3}(a) and (b), $I_c(eV_{G,W})$ exhibits peaks associated with quantum-size oscillations, in agreement with $I_c(W)$ in Fig.~\ref{FigSs2}(b).  These peaks are determined by the properties of the nanoribbon (not the junction) and are therefore independent of $eV_{G,w}$. In contrast, when only the gating inside the junction ($V_{G,w}$) is varied, the corresponding $I_c(V_{G,w})$ has a richer resulting behavior, with a resonance peak around $V_{G,w}=-18$ meV and a {\it double peak} structure between $V_{G,w}=0$ and $30$ meV [see Fig.~\ref{FigSs3}(a)]. In particular, the double peak is clearly observed in $I_c(V_{G,w})$ for $V_{G,W}=0$ [see Fig.~\ref{FigSs3}(c)]. However, the occurrence of these peaks in $I_c(V_{G,w})$ does depend on $V_{G,W}$, in clear opposition to the case of $I_c(V_{G,W})$ whose features are independent of $V_{G,w}$.  For example, as highlighted in Fig.~\ref{FigSs3}(c), the first peak in the double peak structure in $I_c(V_{G,w})$ shifts to higher voltage when the gate voltage outside the junction $eV_{G,W}$ is increased from $0$ to $35$ meV, while the second peak entirely disappears.

To get insight into the origins of the double-peak behavior in the $I_c(eV_{G,w})$, we examine the normal-state density of states in the junction, $n_w$, as a function of the Fermi energy $E_F$, for different gate voltages applied in the junction, as plotted in Fig.~\ref{FigSs4}.  The applied gate voltage results in the potential difference $\Delta U = eV_{G,W}-eV_{G,w}$ between the junction and the rest of the nanoribbon so that a potential barrier (well) is formed in the junction when $\Delta U$ is positive (negative).
When $\Delta U \geqslant 0$, only smooth and broadened peaks are found in the $n_w$, due to the potential barrier in the junction which prevents the localization of electronic states inside the junction, leading to the formation of fewer peaks.  On the other hand, when $\Delta U \leqslant 0$, we find that the peaks do not only become sharper, but also additional peaks appear.  This is due to the fact that the realized potential well inside the junction can support more localized states (analogously to particle in a box problem).

Therefore, we conclude that selective gating enables rich and broadly tunable behavior of the critical current of the quantum-confined Josephson junction. This rich behavior stems from a nontrivial combination of (i) the quantum resonances in either nanoribbon or constriction, determined by geometrical parameters of the sample, and shifted independently by gating outside or inside the junction, respectively, and (ii) the effects due to a potential barrier or potential well realized in the junction, depending on the applied gating inside and outside of the junction.  

\section{Conclusions}\label{sec:5}

In summary, we have detailed the properties of superconducting nanoribbons with a constriction by solving the Bogoliubov-de Gennes equations self-consistently, in the regime where quantum confinement is of crucial importance. The constriction in the nanoribbon is introduced by two adjacent steps in the lateral edge.  

For a long constriction, the interaction between the adjacent step-edges can be neglected as they are separated by large distance. In this case, we reported the effect of a single step-edge on the superconducting order parameter and the local density of states. We found that the shape resonances of the superconducting gap are different and spatially confined inside and outside the constricted area, separated by an abrupt change in the superconducting properties at the step-edge, on a scale of the Fermi wavelength $\lambda_F$, thereby forming a near ideal S-S' junction. This is due to the fact that the step-edge scatters the normal electronic states, especially the ones that are near the band edge, leading to a large number of localized states concentrating on either side of the step. We also note that the superconducting (inverse) proximity effect at the step-edge is featured in the LDOS, but has far less prominent role than the change in the electronic states there. 

When the two step-edges are close to each other, they form a short constriction. In this case, the normal-state electronic properties of the constriction can be viewed as those of a quantum dot with spectral broadening effects. In addition, the short constriction in the nanoribbon forms a quantum point contact, leading to a quantum-confined Josephson junction, with properties tuned via quantum-size effects in and out of the constriction. The critical current of the junction exhibits a step-like behavior as a function of the width of the constriction, and can be also tuned by the width of the nanoribbon outside the junction. Finally, we demonstrated a rather effective and versatile tunability of the junction properties by local as well as global electronic gating.

Taking everything into account, and bearing in mind the number of emergent crystalline 2D superconductors whose lateral geometry can be precisely patterned, we expect that our results will generate further ideas for control of the low-dimensional superconducting condensate and quantum tailoring of much needed superconducting quantum devices such as advanced SQUID probes \cite{Vas_2013,Uri_2016, Ana_2014,Emb_2017}, novel single-photon detectors \cite{Wal_2017}, phase-slip and weak-link junctions \cite{Lom_2018}, or Josephson qubits for second generation quantum technology \cite{Wan_2015, Ste_2010, Yan_2016}.

\section*{Acknowledgments}
This work was supported by the Research Foundation-Flanders (FWO-Vlaanderen), the Special Research Funds of the University of Antwerp (TOPBOF), the Italian MIUR through the PRIN 2015 program (contract No. 2015C5SEJJ001), the MultiSuper network, and the EU-COST NANOCOHYBRI action CA16218.


%

\end{document}